\documentclass{dukecei}

\usepackage{pifont}          
\usepackage{threeparttable}  
\usepackage{enumitem}        
\usepackage{cleveref}        

\title{ZKPU: \underline{Z}ero-\underline{K}nowledge \underline{P}rocessing \underline{U}nit for Heterogeneous Workloads}

\author{
    Jonathan Ku$^{1}$\footnote{Correspondence E-mail: jonathan.ku@duke.edu},
    Changchun Zhou$^{1}$,
    Jiawen Wu$^{2}$,
    Sangouk Jeon$^{3}$,
    Cong Guo$^{1}$,
    Yuzhe Fu$^{1}$,
    Bowen Duan$^{1}$,
    Dongsuk Jeon$^{3}$,
    JV Rajendran$^{2}$,
    Yiran Chen$^{1}$,
    Hai ``Helen'' Li$^{1}$\\[1em]
    \normalsize $^{1}$Duke University 
    \normalsize $^{2}$Texas A\&M University 
    \normalsize $^{3}$Seoul National University
}

\begin{document}

\maketitle
\thispagestyle{firstpagestyle} 

\begin{abstract}
Zero-knowledge proofs (ZKPs) have become a fundamental building block for privacy-preserving computation, blockchain systems, and verifiable machine learning. Modern ZKP protocols span two distinct families: curve-based protocols, which rely on large finite-field arithmetic, and hash-based protocols, which operate over native 64-bit fields. Existing hardware accelerators are specialized for one protocol family because they are built on operand-scanning Montgomery multiplication, resulting in fixed-width datapaths, poor utilization across heterogeneous operand widths, and limited scalability.

This paper presents \emph{ZKPU}, a Zero-Knowledge Processing Unit for heterogeneous ZKP workloads. ZKPU introduces a new compute paradigm by co-designing finely integrated product scanning (FIPS) Montgomery multiplication with a reconfigurable Twin Systolic Fabric (TSF), enabling modular multiplication at arbitrary widths in 64-bit increments. Combined with a fine-grained Deque buffer, on-chip twiddle-factor generator, and memory-efficient MSM preprocessing strategy, ZKPU provides a unified architecture for both curve-based and hash-based protocols. Evaluation using Groth16 and Plonky2 shows that ZKPU achieves up to $144\times$ speedup for NTT, $13.6\times$ for MSM, and approximately $7.2\times$ end-to-end application speedup over state-of-the-art accelerators, while being the first architecture to efficiently support both protocol families on a single reconfigurable hardware platform.
\end{abstract}

\vspace{1em}

\section{Introduction} \label{sec:intro}
\begin{figure}[htbp] 
  \centering
  \includegraphics[width=.7\linewidth]{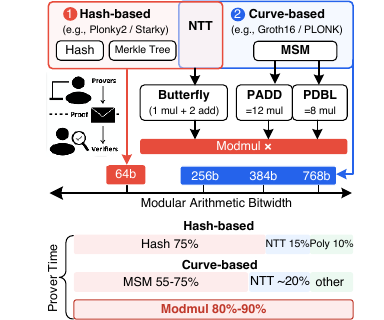}
  \caption{Curve-based and hash-based ZKP protocols rely on distinct computational kernels and arithmetic primitives, resulting in heterogeneous workloads that motivate the design of ZKPU, a unified processing unit for both protocol families.}
  \label{fig:teaser}
\end{figure}

Zero-knowledge proofs (ZKPs)~\cite{zkp} enable a prover to convince a verifier that a statement is true without revealing any information beyond its validity, and have become a fundamental building block for blockchain systems~\cite{zcash,zcoin,survey}, verifiable computation~\cite{zksnark2}, and machine-learning verification\cite{vml,zkcnn,zkml}, etc~\cite{rollup}. Practical ZKP protocols broadly fall into two families. Curve-based protocols, such as Groth16~\cite{groth16} and PLONK~\cite{plonk}, rely on elliptic-curve cryptography (ECC) to generate succinct proofs with fast verification. Their prover execution is dominated by arithmetic-intensive kernels, primarily number-theoretic transforms (NTTs) and multi-scalar multiplication (MSM), over $256\sim768$-bit prime fields. In contrast, hash-based protocols, including Plonky2~\cite{plonky2} and Starky~\cite{starky}, rely on ZKP-friendly hash functions such as Poseidon~\cite{poseidon}, eliminating the need for trusted setup while providing post-quantum security. Their prover execution is dominated by NTTs and Merkle-tree construction, where each hash computation operates over the 64-bit Goldilocks field, resulting in computational characteristics that differ fundamentally from curve-based provers. This is illustrated in \Cref{fig:teaser}.

Despite their algorithmic diversity, the core kernels of ZKP provers reduce to modular arithmetic. NTT, MSM, and Poseidon hash are all dominated by modular addition and multiplication, with modular multiplication accounting for most of the computation. However, operand bitwidths vary by nearly $12\times$, ranging from 64-bit Goldilocks fields to 256/384-bit prime fields (BN254 and BLS12-381) and up to 768-bit fields (MNT4-753).

\begin{table}[htbp]
  \centering
  \caption{Architectural comparison of representative ZKP acceleration works. } \label{tab:config}
  \begin{threeparttable}
    \resizebox{\textwidth}{!}{%
    \begin{tabular}{c|ccccc}
      \toprule
      Work   & Method    & Operand Granularity   & Supported Protocols & Cryptographic Backend & Reconfigurability \\
      \midrule
      PipeZK & CIOS      & 256 or 384 or 768-bit & Groth16 (BN128,BLS12-381,MNT4-753)             & Curve-based           & x                 \\
      LegoZK & CIOS      & 256/384/768-bit       & Groth16 (BN128,BLS12-381,MNT4-753)             & Curve-based           & v                 \\
      UniZK  & Classical & 64-bit                & Plonky2, Starky      & Hash-based            & x                 \\
      ZKPU   & FIPS      & 64*n-bit ($n=1...12$) & Groth16 (BN128,BLS12-381,MNT4-753), Plonky2     & Curve+Hash-based      & v                 \\
      \bottomrule
    \end{tabular}%
    }
  \end{threeparttable}
\end{table}

Prior works as shown in \Cref{tab:config} scale downward: operand-scanning datapaths designed for the maximum supported operand width can only be partitioned into narrower multipliers, reducing utilization for smaller operands while leaving wider operands unsupported. Runtime width flexibility is therefore limited to downward partitioning. For example, LegoZK~\cite{legozk} employs a reconfigurable PE with a fixed 768-bit interface that can be split into 256- or 384-bit lanes, whereas all other designs fix the operand bitwidth at design or synthesis time. To overcome this limitation, we propose \textbf{ZKPU}, a \textbf{Zero-Knowledge Processing Unit for Heterogeneous Workloads}, which scales upward by composing 64-bit processing units into arbitrary widths in 64-bit increments. As a result, a single architecture efficiently supports 64-bit hash-based protocols, $256\sim768$-bit curve-based protocols, and future protocols with even wider operands.

We trace this limitation to a common root cause: existing accelerators are built on operand-scanning Montgomery multiplication (CIOS/FIOS~\cite{scanning}), whose datapath is inherently tied to the operand bitwidth. This coupling fixes the supported width at design time, increases on-chip operand storage, and limits data reuse. To break this dependency, we instead adopt finely integrated product scanning (FIPS)~\cite{scanning}, which operates on 64-bit chunks and naturally composes into modular multipliers of arbitrary widths in 64-bit increments. Although FIPS has long been considered difficult to accelerate because of its irregular operand ordering, we show that these access patterns map efficiently onto a two-dimensional systolic-like architecture, enabling scalable modular arithmetic across heterogeneous ZKP workloads.

We realize this insight with the Twin Systolic Fabric (TSF), a reconfigurable two-dimensional array of 64-bit processing elements that implements FIPS-based modular multiplication. By composing the same 64-bit building blocks, TSF scales seamlessly from native 64-bit hashing to $256\sim768$-bit EC operations without changing the underlying hardware. Sixteen TSF-based ZK Engines, integrated with our proposed fine-grained operand Deque buffer and on-chip twiddle-factor generator, together form \textbf{ZKPU}.

\subsection{Our Contributions}
\setlist{nolistsep}
\begin{itemize}[left=0pt]
\item \textbf{A width-scalable modular multiplication paradigm.} We identify operand-scanning Montgomery multiplication as the root cause of fixed-width ZKP accelerators and introduce a co-designed compute paradigm that combines FIPS with the TSF, enabling modular multiplication at arbitrary widths in 64-bit increments.

\item \textbf{A unified architecture for heterogeneous ZKP workloads.} TSF composes the same 64-bit processing elements into modular multipliers spanning $64\sim768$ bits, efficiently supporting both curve-based and hash-based ZKP protocols while improving utilization and reducing operand movement through systolic data reuse.

\item \textbf{ZKPU and comprehensive evaluation.} We implement ZKPU in Verilog using a TSMC 16nm technology and evaluate it under unified scaling methodology. ZKPU achieves up to $16\times$ speedup over PipeZK and up to $4.3\times$ over LegoZK, while being able to efficiently support both Groth16 and Plonky2 on a single reconfigurable architecture.
\end{itemize}

\subsection{Paper Organization}
The remainder of this paper is organized as follows. \Cref{sec:background} reviews ZKP protocols and the background of Montgomery multiplication. \Cref{sec:arch} presents the proposed ZKPU architecture, including the Twin Systolic Fabric (TSF) and ZK Engine. \Cref{sec:algo} describes the computational characteristics of curve-based and hash-based ZKPs and their mapping onto ZKPU. \Cref{sec:eval} details the implementation methodology and experimental evaluation. \Cref{sec:related} discusses related work on ZKP accelerators, and \Cref{sec:conclusion} concludes the paper.

\section{Background and Motivation} \label{sec:background}

\subsection{Zero-Knowledge Proof Protocols}
ZKPs enable a prover to convince a verifier of a statement's validity without revealing any additional information. They have been widely adopted in blockchain systems, cryptocurrencies~\cite{zcash,zcoin,survey}, verifiable machine learning~\cite{vml}, online auctions~\cite{auction}, and privacy-preserving computation. Among practical constructions, zk-SNARKs~\cite{zksnark1,zksnark2,zksnark3,zksnark4} generate succinct proofs with fast verification, whereas zk-STARKs~\cite{zkstark1,zkstark2,zkstark3} provide transparent setup, post-quantum security, and improved scalability.

From a computational perspective, ZKP protocols can be broadly classified into two families: curve-based and hash-based. Curve-based protocols rely on ECC to generate proofs, while hash-based protocols employ ZKP-friendly hash functions, such as Poseidon~\cite{poseidon}, instead of conventional cryptographic hashes (e.g., SHA-256), which are inefficient for proof generation. Although these protocol families differ in their cryptographic primitives, both ultimately reduce to modular arithmetic and share common computational kernels, motivating a unified hardware architecture.

\begin{figure}[htbp] 
  \centering
  \includegraphics[width=\linewidth]{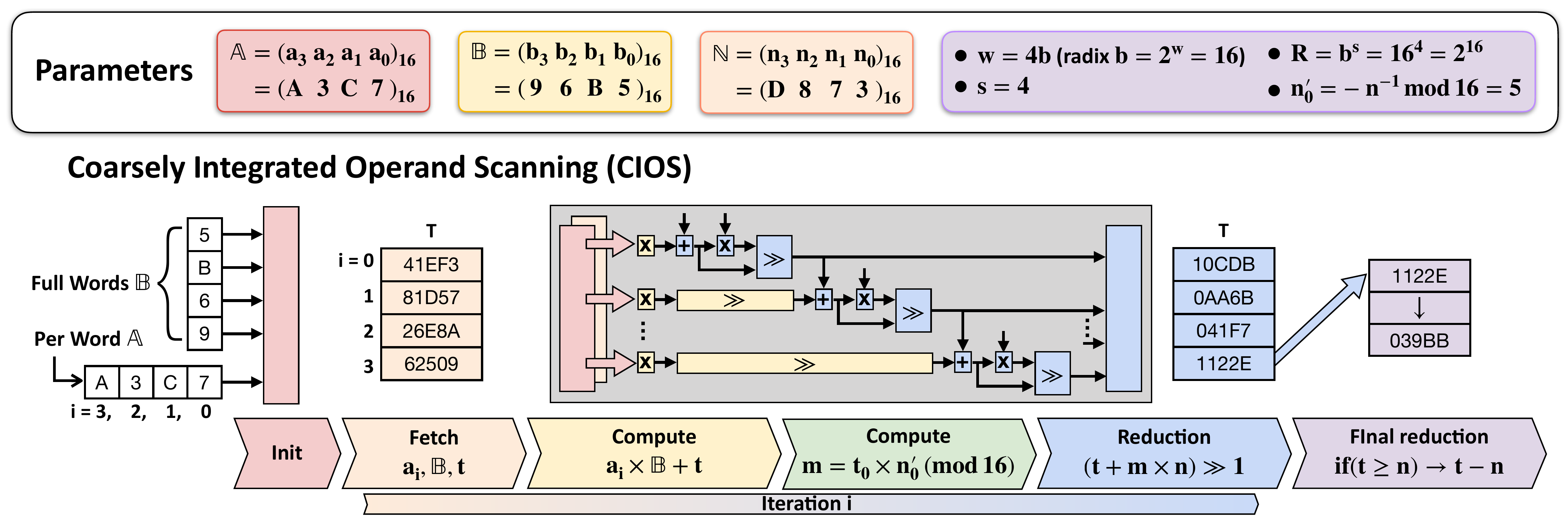}
  \caption{Example for Montgomery multiplication in CIOS method.}
  \label{fig:cios}
\end{figure}
\begin{figure}[htbp] 
  \centering
  \includegraphics[width=\linewidth]{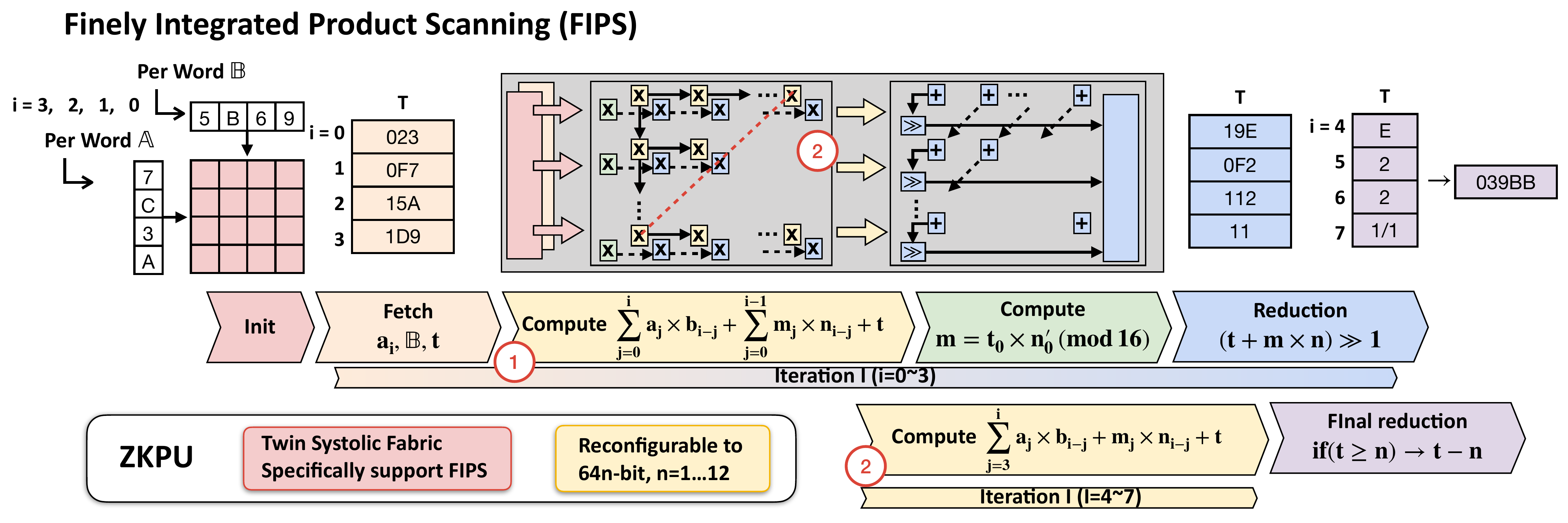}
  \caption{Example for Montgomery multiplication in FIPS method on our ZKPU.}
  \label{fig:fips}
\end{figure}

Modular multiplication is the computational bottleneck in many public-key cryptosystems, including ZKP provers. Among existing reduction algorithms, Montgomery multiplication~\cite{mont} is the de facto choice because it eliminates costly division operations and efficiently supports repeated modular multiplications. To handle large operands ranging from 64 to 768 bits, practical implementations partition operands into multiple machine words and employ multi-word scanning algorithms~\cite{scanning,scanning2,flalm}. Different scanning orders expose distinct dataflows and hardware characteristics. In this work, we focus on operand scanning and finely integrated product scanning (FIPS), and show how co-designing the architecture with FIPS enables efficient, width-scalable modular multiplication.

\begin{figure}[htbp] 
  \centering
  \includegraphics[width=\linewidth]{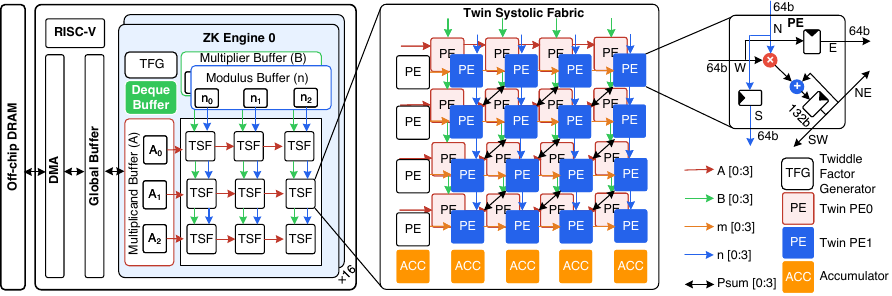}
  \caption{Architecture overview of ZKPU. Left: sixteen ZK
  Engines shared the global buffer, and each engine couples a Deque
  buffer, an on-chip twiddle factor generator, and 9 reconfigurable
  TSFs. Middle: TSF is a two-dimensional array of 64-bit twin PEs,
  where operands $A$, $B$, $m$, and $n$ stream in from the array edges,
  partial sums accumulate along anti-diagonals, and each PE fuses the
  product multiply with the Montgomery reduction multiply. Right: the
  same physical cells regroup as $N{\times}N$ tiles ($N=\text{width}/64$),
  $1{\times}1$ for 64-bit Goldilocks hashing, $4{\times}4$ for 256-bit
  curves, and $12{\times}12$ for 768-bit MNT4-753.}
  \label{fig:zkpu_a}
\end{figure}

\noindent \textbf{Coarsely/Finely Integrated Operand Scanning (CIOS/FIOS).}
CIOS and FIOS are the most widely adopted multi-word Montgomery multiplication algorithms because they interleave modular multiplication with Montgomery reduction, significantly reducing intermediate storage requirements. As illustrated in \Cref{fig:cios}, both algorithms traverse the multiplier $B$ sequentially while keeping the multiplicand $A$ stationary, making them input-oriented scanning methods. For each machine-word of $B$, partial products are generated with every word of $A$. FIOS performs Montgomery reduction immediately after each partial product, whereas CIOS accumulates all partial products for the current word of $B$ before performing a reduction, trading finer granularity for lower reduction overhead. Consequently, hardware implementations typically stream $B$ through the pipeline while storing the full operand $A$ on-chip.

This scan order has been widely adopted in cryptographic accelerators~\cite{pipezk} because of its regular pipeline structure. However, it also introduces three architectural limitations. First, the datapath is inherently tied to the full operand bitwidth since each iteration requires the complete multiplicand $A$, making modular multipliers difficult to reconfigure across different bitwidths. Second, the stationary full-width operand increases on-chip storage and memory bandwidth requirements, especially when multiple modular multipliers execute concurrently. Finally, because computation progresses along a single operand dimension, opportunities for data reuse are limited, reducing hardware utilization for heterogeneous ZKP workloads. 

\noindent \textbf{Finely Integrated Product Scanning (FIPS).}
FIPS adopts a fundamentally different scan order from CIOS/FIOS. Rather than traversing the input operands, it scans the output product from the least- to most-significant word, as shown in \Cref{fig:fips}. Each iteration computes and immediately finalizes one output word after Montgomery reduction, making FIPS an output-oriented scanning algorithm. Consequently, only a single machine word from each operand is required at every step, reducing the working set from full-width operands to fixed 64-bit chunks. This fine granularity naturally enables modular multipliers of arbitrary widths by composing additional 64-bit processing units instead of redesigning the datapath.

Despite these advantages, FIPS has rarely been adopted in hardware because its operand access pattern is irregular. Successive iterations require non-consecutive operand pairs while interleaving partial-product accumulation with Montgomery reduction, making the dataflow difficult to map onto conventional one-dimensional pipelines. The key insight of this work is that these seemingly irregular access patterns become highly regular when mapped onto a two-dimensional systolic-like architecture. This co-design between FIPS and the proposed TSF transforms FIPS from an algorithmic alternative into the foundation of a width-scalable architecture for heterogeneous ZKP workloads.

\noindent \textbf{Coarsely Integrated Hybrid Scanning (CIHS).}
CIHS combines CIOS and FIPS to balance the regular data access of operand scanning with the fine-grained output generation of product scanning. While this hybrid approach can improve the arithmetic efficiency of a standalone modular multiplier, it does not address the architectural limitations discussed above. Supporting both scan orders requires separate datapaths and control logic, increasing hardware complexity and resource overhead. Consequently, CIHS has not been adopted in existing ZKP accelerators, where a single, regular datapath is preferred for high-throughput implementations. In contrast, our work demonstrates that a hardware architecture designed specifically for FIPS can achieve both scalability and efficiency without relying on a hybrid datapath.


\section{ZKPU Architecture} \label{sec:arch}
This section presents the proposed Twin Systolic Fabric (TSF), a reconfigurable architecture for width-scalable modular multiplication. We first describe the TSF and its processing elements (PEs), followed by the Deque buffer and on-chip twiddle-factor generator, which together provide the fine-grained operand supply required by the fabric. These components collectively form a ZK Engine, the fundamental compute-unit of ZKPU. The overall architecture hierarchy is shown in \Cref{fig:zkpu_a}. The number of instantiated ZK Engines is an implementation parameter determined by area and memory constraints.

\begin{figure}[htbp] 
  \centering
  \includegraphics[width=\linewidth]{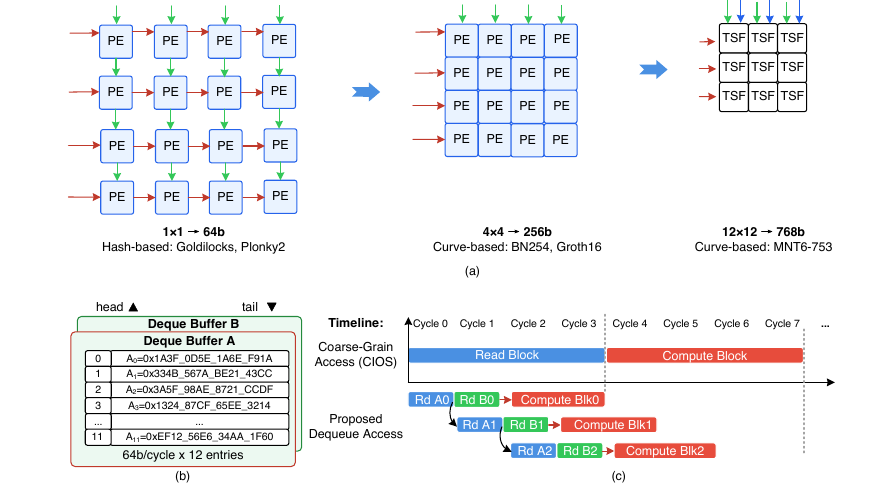}
  \caption{Architecture overview of ZKPU. (a) Reconfigurable PEs configurations supporting 64-, 256-, and 768-bit modular multiplication. (b) Deque buffer enabling fine-grained, bidirectional operand access. (c) Comparison of operand access patterns between conventional CIOS and the proposed FIPS dataflow.}
  \label{fig:zkpu_b}
\end{figure}

\subsection{Twin Systolic Fabric} \label{subsec:tsf}  
Most existing ZKP accelerators adopt CIOS because its input-oriented scan order maps naturally to a simple one-dimensional pipeline. However, this design fundamentally limits scalability across heterogeneous ZKP workloads. First, CIOS requires one operand to remain stationary throughout the computation, tying the datapath to a fixed operand bitwidth. Consequently, supporting different elliptic curves or hash-based protocols requires either separate hardware or specialized reconfigurable designs~\cite{legozk}, which still cannot efficiently span the $64\sim768$-bit operand range. Second, each modular multiplication requires full-width operands to be stored on-chip while one operand is streamed through the pipeline, resulting in substantial storage and memory bandwidth demands. These limitations motivate a new architecture that is both width-scalable and bandwidth-efficient.

To overcome these limitations, we adopt FIPS as the foundation of our modular multiplier. Unlike CIOS, FIPS requires non-consecutive operand pairs at each iteration, making its dataflow difficult to map onto conventional one-dimensional pipelines. Our key insight is that this seemingly irregular access pattern naturally maps onto a two-dimensional systolic-like architecture. Accordingly, we propose TSF, where operands stream horizontally and vertically through the array, allowing both operands to be fully pipelined without requiring either to remain stationary. Furthermore, each processing element operates on a 64-bit word, enabling the same fabric to serve as a native 64-bit engine for hash-based ZKPs or to compose seamlessly into larger modular multipliers for $256\sim768$-bit curve-based protocols.

Our TSF is composed of a $4\times4$ array of 64-bit processing elements (PEs), where neighboring PEs cooperate as twin PEs to implement FIPS-based modular multiplication. A single TSF therefore realizes a 256-bit modular multiplier, with the detailed dataflow illustrated in \Cref{fig:zkpu_b}. To support larger curves, multiple TSFs are seamlessly composed into a $12\times12$ PE array, enabling 768-bit modular multiplication without modifying the underlying hardware. The same 64-bit PEs also natively execute hash-based kernels, allowing a unified architecture to efficiently support both curve-based and hash-based ZKP workloads.

\begin{figure}[htbp] 
  \centering
  \includegraphics[width=\linewidth]{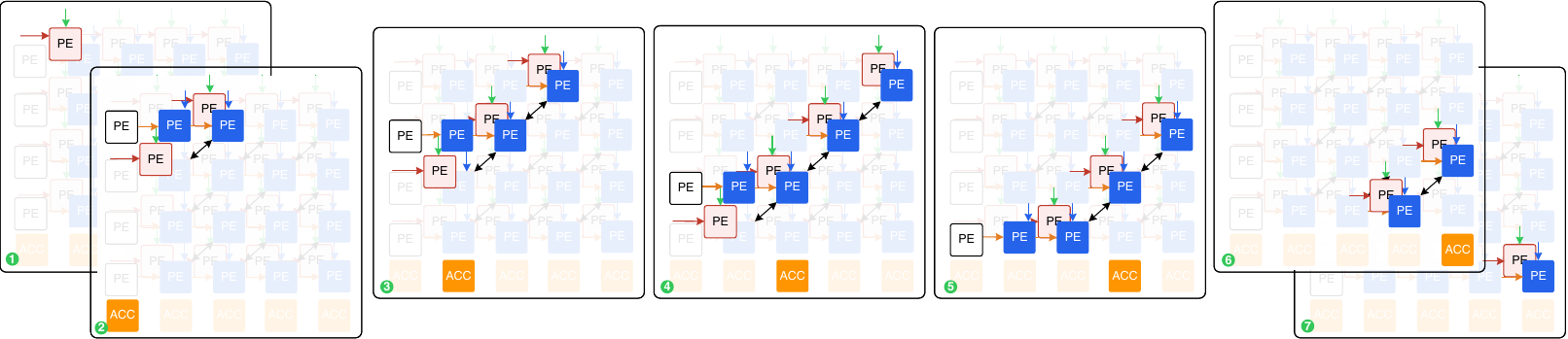}
  \caption{FIPS Montgomery Multiplication dataflow under a 4x4 TSF. 
  \ding{182} Start regular multiplication on $a_0b_0$, 
  \ding{183} Continue regular multiplication on $a_0b_1+a_1b_0$, start reduction term $m_0$ and bypass to compute partial sum $m_0n_0+m_0n_1$,
  \ding{184} Continue regular multiplication on $a_0b_2+a_1b_1+a_2b_0$, start reduction term $m_1$ and bypass to compute partial sum $m_1n_0 + m_1n_1$ and continue partial sum $m_0n_2$
  \ding{185} Continue regular multiplication on $a_0b_3+a_1b_2+a_2b_1+a_3b_0$, start reduction term $m_2$ and bypass to compute partial sum $m_2n_0 + m_2n_1$ and continue partial sum $m_1n_2+m_0n_3$
  \ding{186} Continue regular multiplication on $a_1b_3+a_2b_2+a_3b_1$, start reduction term $m_3$ and bypass to compute partial sum $m_3n_0 + m_3n_1$ and continue partial sum $m_2n_2+m_1n_3$
  \ding{187} Continue regular multiplication on $a_2b_3+a_3b_2$, and continue partial sum $m_2n_3+m_3n_2$
  \ding{188} Finish regular multiplication on $a_3b_3$, and finish partial sum $m_3n_3$
  } 
  \label{fig:illus}
\end{figure}

\subsection{Deque buffer} \label{subsec:deque}
To fully exploit the fine-grained parallelism of TSF, the memory system must provide operands at the same 64-bit granularity as the processing elements. Conventional operand buffers are optimized for full-width accesses, making them inefficient for TSF, which consumes only a 64-bit word from each operand per cycle. We therefore propose a double-ended (Deque) buffer that streams operands directly from the least-significant word, eliminating the need to preload entire operands on-chip. This fine-grained access allows more modular multiplications to execute concurrently under the same memory bandwidth, significantly improving throughput for curve-based ZKPs as shown in \Cref{fig:zkpu_b}. For hash-based ZKPs, whose operands are naturally 64 bits wide, the same Deque buffer is used without modification, enabling a unified memory architecture across heterogeneous workloads.

The Deque buffer is also designed to accelerate multi-scalar multiplication (MSM), one of the dominant kernels in curve-based ZKPs. Our MSM optimization performs lightweight preprocessing to reorganize points before Pippenger's algorithm, reducing redundant elliptic-curve operations. The algorithmic details are presented in \Cref{sec:algo}; here we focus on the supporting hardware. MSM datasets typically range from $2^{14}$ to $2^{24}$ points, resulting in memory footprints from several megabytes to hundreds of megabytes. Efficient data management is therefore critical. The Deque buffer supports insertion of newly fetched points while allowing removal from both ends. As points are consumed during preprocessing, they become unnecessary and can be discarded immediately, whereas less relevant points at the opposite end are evicted when the buffer reaches capacity. This double-ended organization minimizes unnecessary data movement while sustaining a continuous stream of points for MSM acceleration.

\subsection{ZK Engine} \label{subsec:zke}
The ZK Engine integrates the TSF, Deque buffer, and an on-chip twiddle-factor generator into a self-contained compute unit for ZKP kernels. Since NTT is a fundamental operation in both curve-based and hash-based protocols, twiddle factors are generated on-chip to eliminate frequent off-chip accesses, significantly reducing memory bandwidth with minimal area overhead. By combining fine-grained operand streaming with local twiddle generation, each ZK Engine operates largely independently, allowing multiple engines to execute concurrently. Our ZKPU instantiates 16 ZK Engines connected to off-chip DRAM, although the number of engines is an implementation parameter determined by the target area and memory budget.

\subsection{FIPS Montgomery Multiplication Illustration} \label{subsec:illus}
An example of the TSF dataflow is illustrated in \Cref{fig:illus}. Computation begins at PE(0,0), which generates the least-significant partial product. As operands stream horizontally and vertically through the array, successive PEs compute increasingly significant partial products. Unlike a conventional systolic array, each twin PE simultaneously computes the corresponding Montgomery reduction term. The generated reduction value is immediately forwarded to neighboring twin PEs, allowing modular multiplication and Montgomery reduction to proceed concurrently within the same dataflow. After the initial pipeline fill, both computations are fully overlapped, maximizing processing-element utilization.

The output dataflow follows the anti-diagonal traversal of FIPS. Partial products and reduction terms produced by each PE are accumulated along anti-diagonals to form the intermediate results. During the early stages, the accumulated values generate the Montgomery reduction terms required for subsequent iterations. Once the reduction stream is established, each anti-diagonal produces one finalized output word, with a final conditional reduction applied only if the accumulated result exceeds the modulus. This output-oriented dataflow naturally matches the FIPS algorithm while preserving the regular communication pattern of the two-dimensional systolic fabric.

\section{Accelerating ZKP Protocols} \label{sec:algo}
This section describes how the proposed ZKPU accelerates the fundamental kernels of both curve-based and hash-based ZKP protocols. \Cref{tab:zkp} shows the feature and differences for both families of ZKP protocols. Since these kernels ultimately reduce to modular arithmetic, ZKPU adopts modular multiplication as the primary building block and implements it using the co-designed FIPS-based TSF. The same 64-bit processing elements seamlessly compose into wider modular multipliers, enabling efficient execution across operand widths ranging from 64 to 768 bits. Building upon this foundation, we show how the major ZKP kernels are mapped onto ZKPU to support heterogeneous protocols within a unified architecture.

\begin{table}[htbp]
  \centering
  \caption{Comparison of features, pros \& cons in curve- and hash-based ZKP protocols.} \label{tab:zkp}
  \begin{tabular}{l|ll}
    \toprule
    Feature            & Curve-based ZKPs & Hash-based ZKPs \\
    \midrule
    Primitives   & NTT \& MSM & NTT \& Hash \& \\
    Bottleneck   & MSM Latency & Memory Bandwidth \\
    Finite field & Large (e.g., $256\sim768$-bit) & Small (e.g., 64-bit) \\
    Proof size         & Small (hundreds of bytes) & Large (KB to MB) \\
    Verification & Faster & Scalable \\
    Trusted Setup      & Required & Transparent (None) \\
    Quantum resilient  & No (Vulnerable to Shor's) & Yes \\
    \bottomrule
  \end{tabular}
\end{table}

\subsection{Curve-based Protocols} \label{subsec:curve}
Curve-based ZKP protocols, such as Groth16~\cite{groth16} and PLONK~\cite{plonk}, generate proofs through polynomial arithmetic and EC operations. The proving process first translates the computation into a constraint system (e.g., R1CS), from which the corresponding polynomial representations are constructed. Polynomial evaluation and interpolation are efficiently performed using number-theoretic transforms (NTTs), whose sizes typically range from $2^{14}$ to $2^{24}$. The resulting polynomial coefficients are then used in multi-scalar multiplication (MSM), where predetermined EC points are combined using algorithms such as Pippenger's~\cite{pippenger}. Depending on the underlying curve, MSM operates over either the G1 or G2 group, with G2 arithmetic incurring approximately four times the computational cost of G1 due to its more complex finite-field operations. Consequently, NTT and MSM dominate the prover execution time and constitute the primary acceleration targets for curve-based ZKP protocols.

\noindent \textbf{Number Theoretic Transform (NTT).} Similar to prior accelerators~\cite{pipezk,legozk}, ZKPU adopts the 4-step NTT algorithm to recursively decompose large NTTs into smaller kernels, enabling support for arbitrary transform sizes while matching the hardware tile dimensions. Unlike previous designs, however, the underlying compute fabric is not restricted to a fixed operand width. Since TSF composes 64-bit processing elements into modular multipliers of arbitrary widths, the same NTT kernel can be executed efficiently over both 64-bit Goldilocks fields and 256--768-bit prime fields without modifying the hardware. Consequently, the same NTT architecture is shared across both hash-based and curve-based ZKP protocols.

The fine-grained dataflow of FIPS further reduces memory traffic during NTT execution. Conventional operand-scanning accelerators preload an entire operand before computation, requiring full-width memory accesses for every butterfly operation. In contrast, TSF consumes only the least-significant 64-bit word of each operand at every cycle, allowing operands to be streamed directly from the Deque buffer. This finer-grained access increases memory concurrency and improves bandwidth utilization, particularly for large curve-based fields.

\begin{figure}[htbp] 
  \centering
  \includegraphics[width=.7\linewidth]{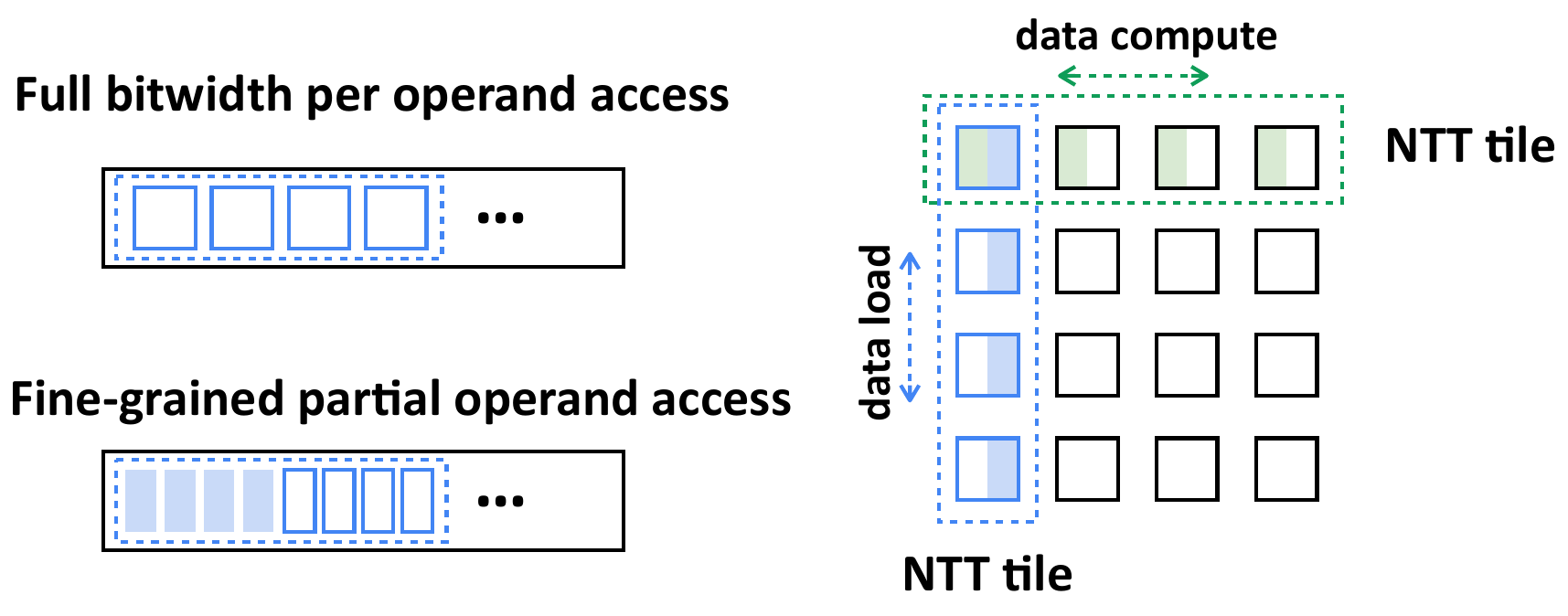}
  \caption{Recursive decomposition and hardware mapping of variable-size NTT kernels on ZKPU.}
  \label{fig:ntt}
\end{figure}

Each ZK Engine directly supports NTT kernels of up to $2^{10}$ points. Larger transforms are recursively decomposed using the 4-step algorithm, while multiple ZK Engines execute independent kernels in parallel. For 64-bit hash-based protocols, all engines operate concurrently to maximize throughput. For wider curve-based fields, larger modular multipliers are constructed by composing multiple TSFs, and the remaining butterfly operations are executed through serialized kernel scheduling similar to LegoZK~\cite{legozk}. This co-design of algorithm and architecture enables ZKPU to efficiently support a wide range of NTT sizes and operand widths, as illustrated in \Cref{fig:ntt}.

\begin{figure}[htbp] 
  \centering
  \includegraphics[width=.7\linewidth]{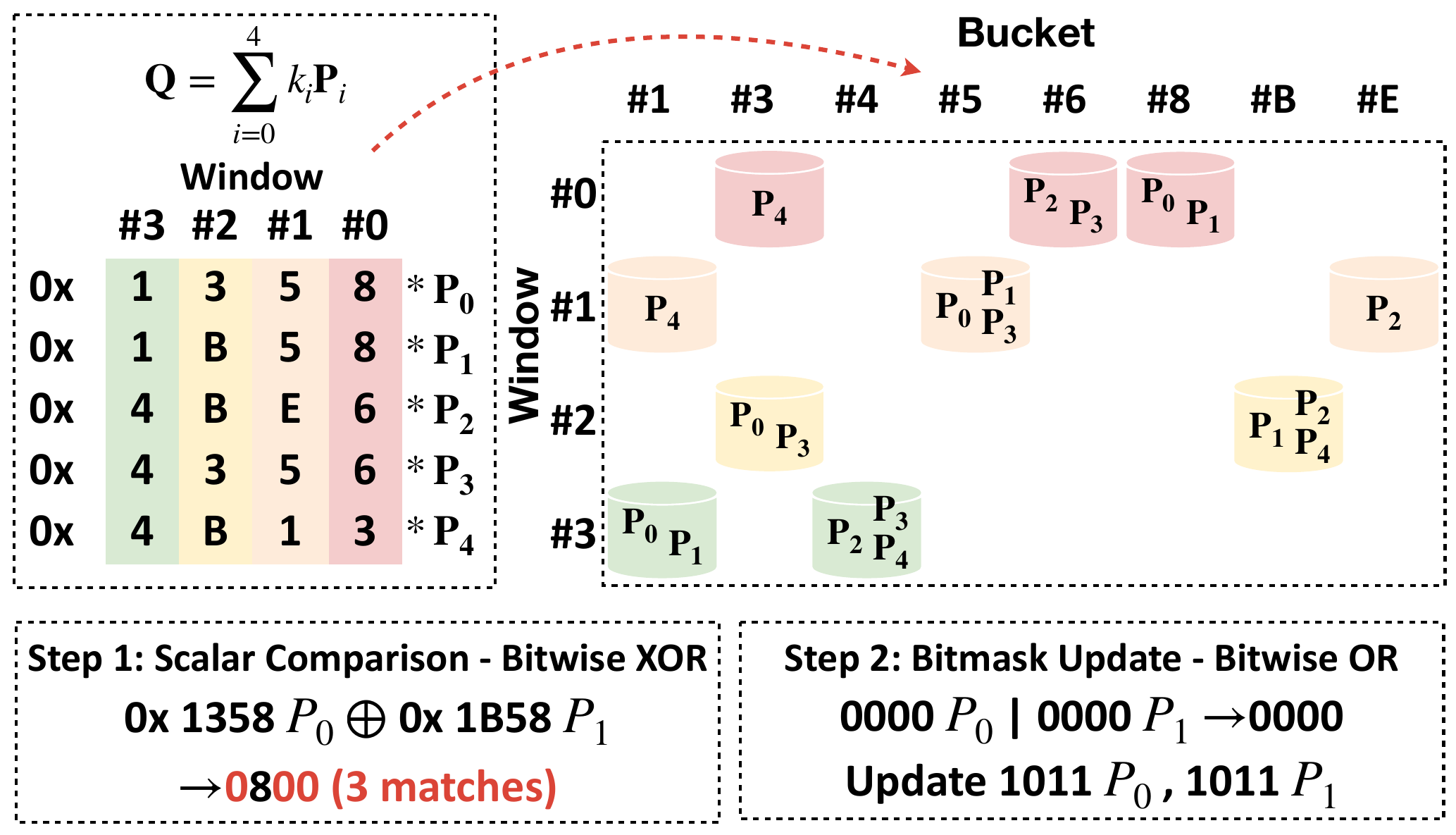}
  \caption{Our strategy for MSM based on Pippenger's algorithm combining points under identical buckets in different windows. The example illustrated used 6 points and }
  \label{fig:pippenger}
\end{figure}

\noindent \textbf{Multi-Scalar Multiplication (MSM).} Multi-scalar multiplication (MSM) is the dominant kernel in curve-based ZKP provers, accounting for more than 80\% of the total proving time~\cite{gzkp}. MSM computes the weighted sum of elliptic-curve points, where each point is multiplied by a scalar through point multiplication (PMUL). PMUL is further decomposed into point additions (PADDs) and point doublings (PDBLs), both of which are implemented using modular arithmetic. To reduce the cost of EC operations, our implementation adopts mixed-addition in Jacobian coordinates~\cite{mixed}, which minimizes the number of field inversions during PADD.

The state-of-the-art algorithm for MSM is Pippenger's algorithm~\cite{pippenger}, which partitions scalars into fixed-size windows and groups elliptic-curve points into buckets according to the window values. Instead of performing each scalar multiplication independently, points assigned to the same bucket are accumulated first, substantially reducing the number of PADD operations. Numerous optimizations have been proposed, including non-adjacent form (NAF)~\cite{naf} encoding to reduce the number of non-zero buckets and multi-window scheduling to improve memory locality and bandwidth utilization~\cite{legozk}.

Building upon Pippenger's algorithm, ZKPU introduces a memory-efficient bucket management strategy tailored to the proposed Deque buffer. Instead of treating bucket construction solely as an algorithmic optimization, our design co-optimizes the preprocessing algorithm and hardware organization to reduce unnecessary data movement while maintaining a continuous stream of points to the compute fabric. As described in the following, this co-design improves memory efficiency without modifying the underlying MSM computation.

Our approach further exploits redundancy in Pippenger's algorithm by identifying identical bucket assignments across different scalar windows, rather than processing each window independently. As illustrated in \Cref{fig:pippenger}, this preprocessing step merges points belonging to identical buckets across multiple windows before bucket accumulation, reducing redundant PADD operations while points are streamed from off-chip memory. Unlike conventional Pippenger, which groups points only within the same window, our method performs bucket matching across windows to improve arithmetic efficiency without modifying the underlying MSM algorithm.

The state-of-the-art algorithm for MSM is Pippenger's algorithm~\cite{pippenger}, which partitions scalars into fixed-size windows and groups EC points into buckets according to the window values. We adopt an 8-bit window, a configuration that provides a good balance between bucket count and memory overhead. In \Cref{sec:eval}, we show the tradeoff for required buffer and bandwidth with point merging redundancy. Instead of performing each scalar multiplication independently, points assigned to the same bucket are accumulated first, substantially reducing the number of PADD operations. Numerous optimizations have been proposed, including non-adjacent form (NAF) encoding to reduce the number of non-zero buckets and multi-window scheduling to improve memory locality and bandwidth utilization~\cite{legozk}. 

To maximize the benefit of our bucket-merging strategy while minimizing off-chip memory traffic, we propose a double-ended queue (Deque) policy tailored for MSM preprocessing. The policy maintains scalars and their associated points on-chip only while they remain likely to participate in future bucket matches. Scalars are inserted from the back of the Deque as they are fetched from memory, while preprocessing proceeds from the front by comparing each scalar with the remaining entries. Once a scalar has completed all possible comparisons, it is immediately removed from the front. Conversely, when the buffer reaches capacity, the most recently inserted entries at the back are the least likely to be matched in the near future and are preferentially evicted to accommodate newly fetched scalars. This double-ended policy extends the effective lifetime of useful data on-chip, reducing redundant memory accesses and improving memory bandwidth utilization during our MSM strategy.

\subsection{Hash-based Protocols}
Hash-based ZKP protocols, such as Plonky2 and Starky, eliminate the trusted setup required by curve-based protocols and instead rely on polynomial commitments and Merkle-tree construction to generate proofs. Similar to curve-based protocols, polynomial operations are accelerated using NTTs, accounting for approximately 15\% of the prover execution time. The remaining computation is dominated by Merkle-tree construction, where each node is computed using a ZKP-friendly hash function. Poseidon hash~\cite{poseidon} has become the de facto choice because it is specifically designed for efficient arithmetic over finite fields, making it significantly more suitable for proof generation than conventional cryptographic hashes. Consequently, Poseidon hashing accounts for roughly 70\% of the prover runtime and, together with NTT, forms the primary acceleration target for hash-based ZKP protocols.

\noindent \textbf{Poseidon Hash.}
Poseidon is a ZKP-friendly cryptographic hash function designed for efficient computation over finite fields. Unlike conventional hash functions, such as SHA-256, which rely heavily on bitwise operations, rotations, and shifts, Poseidon is constructed entirely from finite-field arithmetic. This significantly reduces the arithmetic circuit complexity required for proof generation, making Poseidon the hash function for modern hash-based ZKP protocols.

During proof generation, Poseidon is primarily used for Merkle-tree construction, which dominates the prover execution time. In Plonky2, polynomial evaluations are committed by constructing a Merkle tree over the 64-bit Goldilocks field, where each internal node is computed using the Poseidon hash function. The resulting Merkle root serves as the polynomial commitment. Consequently, efficient acceleration of Poseidon directly translates to higher prover throughput.

Each Poseidon permutation consists of three operations: adding round constants, applying the S-Box, and multiplying by the Maximum Distance Separable (MDS) matrix. Plonky2 uses 8 full rounds and 22 partial rounds. In a full round, all 12 state elements are updated through the S-Box and MDS transformation. In contrast, a partial round applies the S-Box to only a single state element while updating the remaining states through a sparse MDS matrix. This optimization reduces the MDS computation from a dense $t \times t$ matrix-vector multiplication to a sparse formulation, decreasing the arithmetic complexity from $t^2$ to $2t-1$ operations ($t=12$), thereby significantly improving prover performance. 

UniZK~\cite{unizk} demonstrated that a single Poseidon round can be efficiently mapped onto a $12\times12$ systolic array, corresponding to the 12-state permutation used in Plonky2. ZKPU naturally supports the same computation without requiring a dedicated accelerator. A fully composed TSF forms a $12\times12$ PE array, where each PE performs native 64-bit modular arithmetic over the Goldilocks field. Consequently, the existing TSF can execute Poseidon rounds directly with only minor control logic to switch between modular multiplication and Poseidon execution modes, enabling efficient support for both curve-based and hash-based kernels on a unified architecture.

\section{Methodology \& Evaluation} \label{sec:eval}

\begin{table}[htbp]
  \centering
  \caption{Area and power breakdown of previous works.} \label{tab:zkpu}
  \begin{threeparttable}
    \begin{tabular}{cc|cc}
      \toprule
      Work   & Technology Node & Area ($mm^2$) & Power (W) \\
      \midrule
      ZKPU   & 16nm & 24.93    & 11.95     \\
      \midrule
      PipeZK & 28nm & 52.91    & 7.04      \\
      LegoZK & 28nm & 68.79    & 11.39     \\
      UniZK  & 7nm  & 57.8     & 96.4      \\
      \bottomrule
    \end{tabular}
  \end{threeparttable}
\end{table}


\begin{figure}[htbp] 
  \centering
  \includegraphics[width=\linewidth]{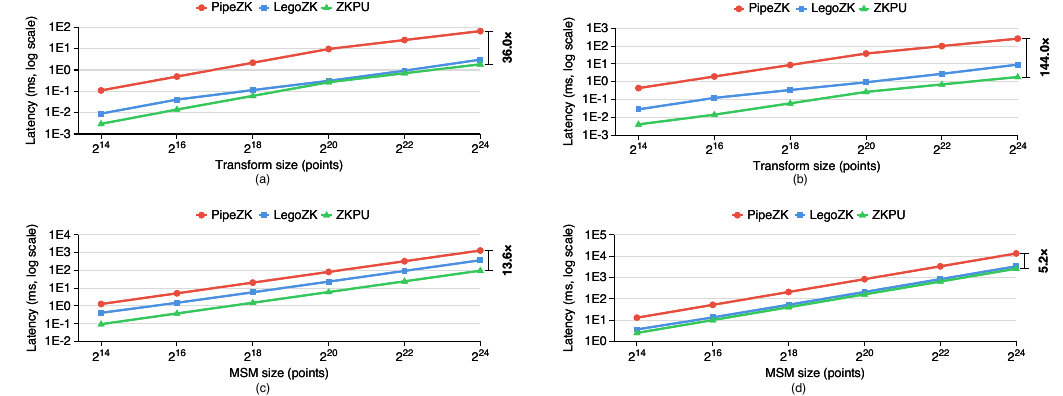}
  \caption{Core kernels in curve-based ZKP protocols: NTT and MSM kernel latency of ZKPU compared with previous works under 300MHz. (a,b) NTT latency vs. transform size at 256-, and 768-bit operand widths. (c,d) MSM latency vs. number of points at 256-, and 768-bit operand widths.}
  \label{fig:latency}
\end{figure}

\subsection{Methodology}
\noindent \textbf{Architecture Implementation.}
To evaluate ZKPU, we develop a cycle-accurate simulator that models the core computational kernels of both curve-based and hash-based ZKP protocols. The simulator maps modular arithmetic operations onto the proposed architecture and evaluates Groth16 and Plonky2 as representative protocols. The key hardware components of ZKPU are implemented in Verilog and synthesized using Synopsys Design Compiler with the TSMC 16nm technology PDK at 300MHz to obtain area and power estimates.

For comparison, we reproduce representative ZKP accelerators, including PipeZK and LegoZK for curve-based protocols and UniZK for hash-based protocols. Since these designs were originally reported under different technology nodes and implementation constraints, we normalize the results using DeepScaleTool~\cite{deepscaletool} and compare all architectures under unified scaling methodology. We further report the scaled latency to jointly evaluate the impact of technology node and area.

\begin{figure}[htbp] 
  \centering
  \includegraphics[width=.7\linewidth]{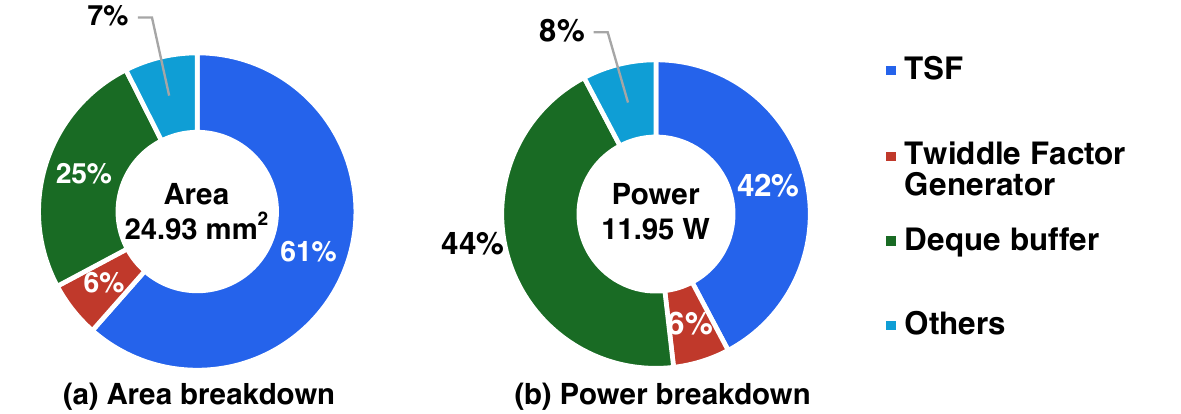}
  \caption{Area and power breakdown of ZKPU in 16nm. The PEs within TSFs dominate in area while both TSF and Deque buffer have similar contribution to power.}
  \label{fig:breakdown}
\end{figure}

\noindent\textbf{Area and Power Breakdown.}
\Cref{tab:zkpu} compares the implementation cost of ZKPU with prior ZKP accelerators under the same 16nm technology node. Although ZKPU targets both curve-based and hash-based protocols within a unified architecture, its area and power remain competitive with specialized accelerators. Compared with PipeZK, LegoZK, and UniZK, ZKPU achieves comparable implementation cost while providing substantially greater flexibility through its reconfigurable TSF and unified hardware support. These results demonstrate that heterogeneous protocol support can be achieved with modest hardware overhead rather than requiring separate accelerators for different ZKP families.

\Cref{fig:breakdown} shows the area and power breakdown of ZKPU. The Twin Systolic Fabric (TSF) occupies 61\% of the total chip area, reflecting its role as the primary compute engine. The Deque buffer accounts for 25\% of the area, while the twiddle-factor generator and other supporting logic contribute only 6\% and 8\%, respectively. In terms of power, TSF and the Deque buffer dominate the consumption at 42\% and 44\%, indicating that both computation and fine-grained operand movement contribute significantly to the overall energy. The twiddle-factor generator incurs only 6\% of the total power, demonstrating that on-chip twiddle generation provides bandwidth savings with minimal hardware overhead.

\begin{table}[htbp]
  \centering
  \caption{Benchmark Characteristics.} \label{tab:benchmark}
  \begin{tabular}{lll|ll}
    \toprule
    Protocol & Application & Size  & Application & Size \\
    \midrule
    Groth16  & AES         & 16k   & SHA         & 32k  \\
             & RSA-Enc     & 96k   & RSA-SHA     & 128k \\
             & Merkle Tree & 288k  & Auction     & 544k \\
    \midrule
    Plonky2  & Factorial   & 1M    & Fibonacci   & 32k  \\ 
    \bottomrule
  \end{tabular}
\end{table}

\noindent \textbf{Applications and Workloads.}
We evaluate ZKPU using representative applications from both curve-based and hash-based ZKP protocols. For curve-based ZKPs, we adopt the same Groth16 benchmark suite used by prior accelerators~\cite{pipezk,legozk}, including \textbf{AES}, \textbf{SHA}, \textbf{RSA-Enc}, \textbf{RSA-SHA}, \textbf{Merkle Tree}, and \textbf{Auction}. Following the evaluation methodology of previous work, we use the same circuit configurations to enable direct comparison. The application characteristics are summarized in \Cref{tab:benchmark}.

For hash-based ZKPs, we evaluate Plonky2 using its standard example applications, including \textbf{Factorial} and \textbf{Fibonacci}. These benchmarks exercise the complete proving pipeline, including NTT, Poseidon hashing, and Merkle-tree construction, providing representative workloads for modern hash-based ZKP systems. The corresponding workload characteristics are also summarized in \Cref{tab:benchmark}.

\begin{figure}[htbp] 
  \centering
  \begin{subfigure}[t]{.75\linewidth}
    \centering
    \includegraphics[width=\linewidth]{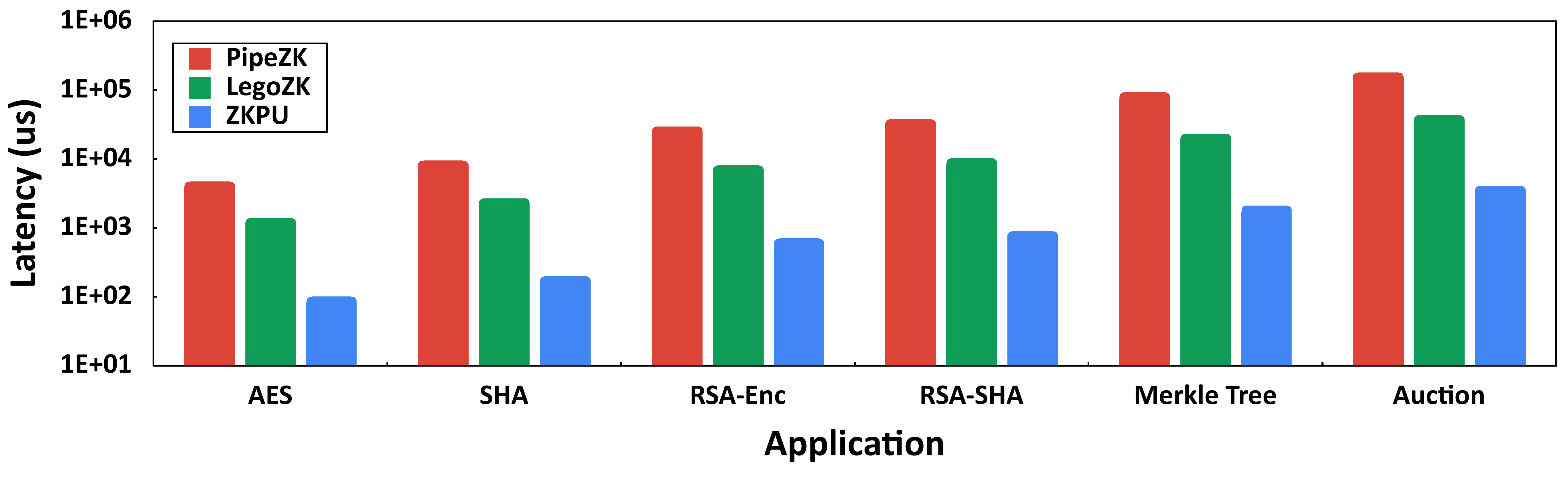}
    \caption{Scaled latency for Groth16 under BN254 curve.}
  \label{fig:groth16_app_256b}
  \end{subfigure}%
  \hfill
  \begin{subfigure}[t]{.75\linewidth}
    \centering
    \includegraphics[width=\linewidth]{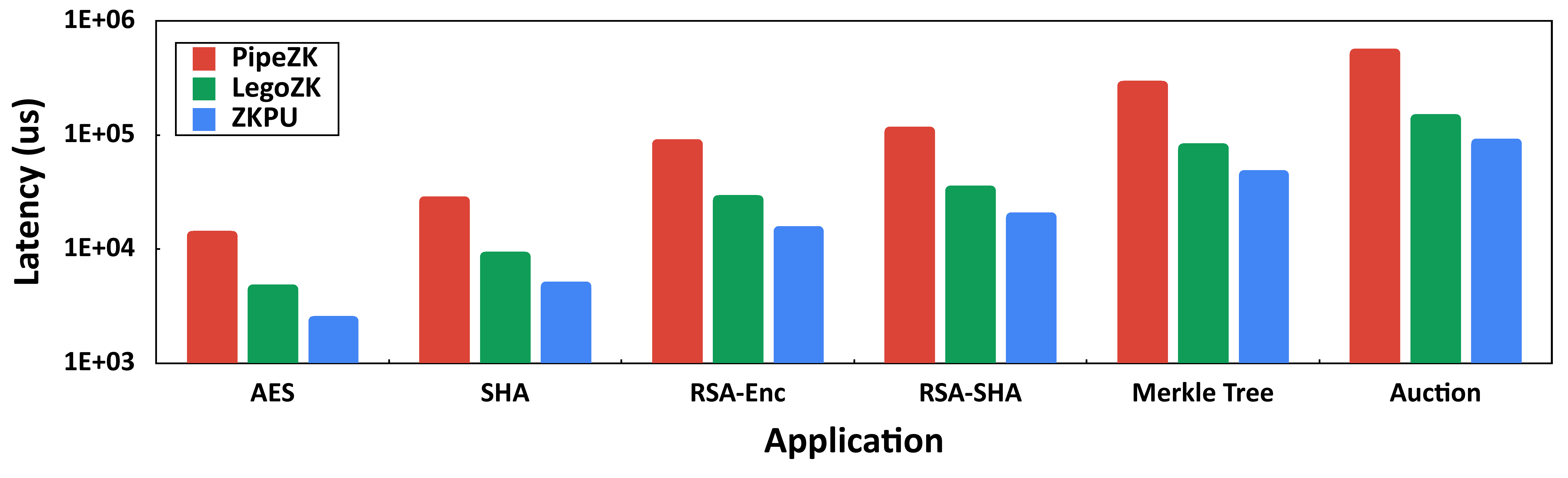}
    \caption{Scaled latency for Groth16 under MNT4-753 curve.}
  \label{fig:groth16_app_768b}
  \end{subfigure}%
  \hfill
  \begin{subfigure}[t]{.245\linewidth}
    \centering
    \includegraphics[width=\linewidth]{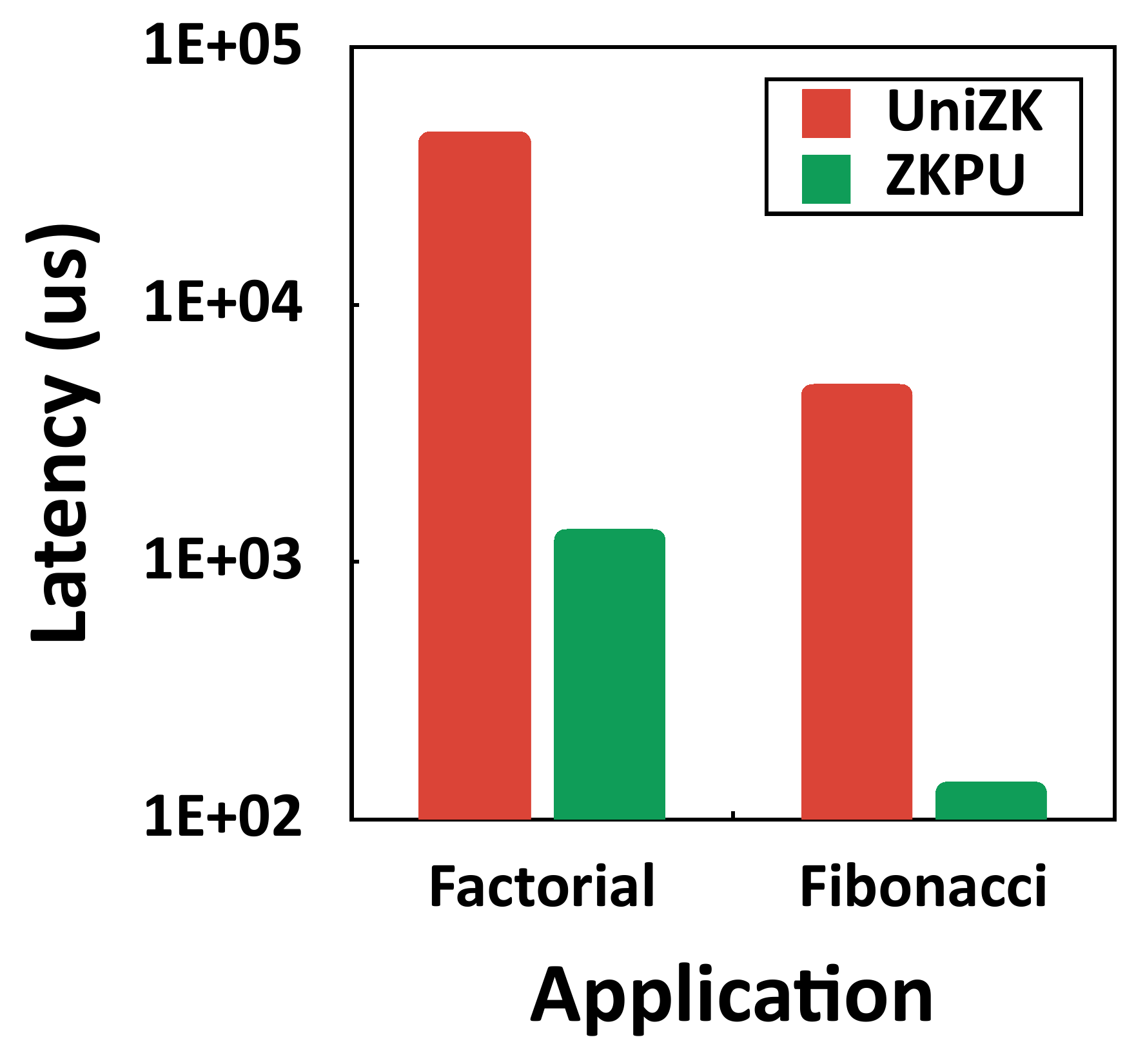}
    \caption{Scaled latency for Plonky2.}
  \label{fig:plonky2_app}
  \end{subfigure}%
  \caption{Scaled latency for different workloads in Groth16 and Plonky2.}
  \label{fig:app}
\end{figure}

\subsection{Performance Evaluation.}
\noindent \textbf{NTT performance.} 
\Cref{fig:latency} compares the NTT latency of ZKPU with PipeZK and LegoZK for both 256-bit and 768-bit modular arithmetic. Across all transform sizes, ZKPU consistently achieves the lowest latency. For 256-bit NTTs, ZKPU provides an average speedup of $36.0\times$ over PipeZK and $2.0\times$ over LegoZK. At 768 bits, the average speedup further increases to $144\times$ and $5.6\times$, respectively.

The speedup over PipeZK remains nearly constant at 256 bits and increases substantially at 768 bits, demonstrating that ZKPU scales efficiently with operand width. In contrast, the fixed-width operand-scanning datapath of prior designs incurs increasing overhead as the operand bitwidth grows. Compared with LegoZK, ZKPU consistently outperforms across all transform sizes, validating the effectiveness of the proposed TSF and fine-grained operand access.

\noindent \textbf{MSM performance.}
\Cref{fig:latency} compares the MSM latency of ZKPU with PipeZK and LegoZK for both 256-bit and 768-bit arithmetic. Across all evaluated MSM sizes, ZKPU exhibits near-linear scalability while consistently achieving the lowest execution latency. At 256 bits, ZKPU provides an average speedup of $13.6\times$ over PipeZK and $3.9\times$ over LegoZK. For 768-bit MSM, ZKPU maintains an average speedup of $5.2\times$ and $1.3\times$, respectively.

The nearly constant speedup across all MSM sizes demonstrates that the proposed preprocessing strategy of Pippenger's algorithm and Deque-buffer design scale efficiently with increasing workloads. With dynamic window size used in LegoZK that would greatly impact the memory bandwidth due to the exponential increase in bucket storage, our strategy is able to find redundancy and thus memory-friendly. Compared with prior accelerators, ZKPU consistently improves MSM performance while supporting both curve-based and hash-based ZKP protocols within a unified architecture.

\noindent \textbf{Groth16 evaluation.}
\Cref{fig:app} compares the end-to-end proving latency of ZKPU with PipeZK and LegoZK across representative Groth16 applications under both 256-bit (BN254) and 768-bit (MNT4-753) curves. ZKPU consistently achieves the lowest latency for all benchmarks. The performance improvement is preserved across both operand widths, demonstrating that ZKPU efficiently supports heterogeneous elliptic curves using the same reconfigurable architecture.

The latency reduction is attributed to the co-design of our architecture and memory policy, which accelerates the dominant modular arithmetic in NTT and MSM while maintaining high utilization across different operand widths. With the Deque buffer for fine-grained operand access, ZKPU sustains higher arithmetic throughput than prior accelerators without sacrificing architectural flexibility.

\noindent \textbf{Plonky2 evaluation.} 
\Cref{fig:app} compares the end-to-end proving latency of ZKPU with UniZK using benchmarks provided by the Plonky2 framework. ZKPU consistently outperforms UniZK, achieving an average speedup of approximately $9.5\times$ across both applications. The results demonstrate that ZKPU efficiently accelerates the dominant kernels of hash-based ZKPs, including NTT and Poseidon hash.

The performance improvement stems from the native support for 64-bit Goldilocks arithmetic, allowing TSF to execute Poseidon and NTT directly without hardware reorganization. ZKPU provides substantially higher throughput while retaining the same unified architecture used for curve-based protocols.

\begin{figure}[htbp] 
  \centering
  \includegraphics[width=.7\linewidth]{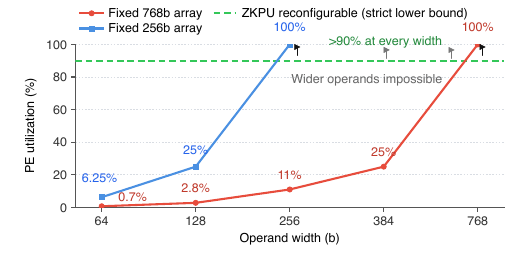}
  \caption{PE utilization versus operand bitwidth for fix-bitwidth implementation and our ZKPU.}
  \label{fig:utilization}
\end{figure}
\begin{figure}[htbp] 
  \centering
  \includegraphics[width=.7\linewidth]{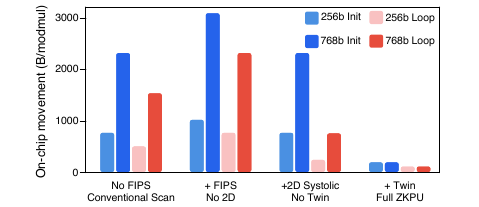}
  \caption{Ablation study of on-chip data movement per modular multiplication with traditional CIOS, FIPS method, simple systolic and our ZKPU. (under analytical 256 and 768-bit). }
  \label{fig:movement}
\end{figure}

\noindent\textbf{PE Utilization.}
\Cref{fig:utilization} compares the PE utilization of ZKPU with conventional fixed-width modular multipliers. A fixed-width 256-bit array achieves full utilization only for 256-bit operands, dropping to 25\% and 6.25\% for 128- and 64-bit operands, respectively, while being unable to support wider operands. Likewise, a fixed 768-bit array is highly underutilized for smaller operands, achieving only 25\%, 11\%, and 0.7\% utilization for 384-, 256-, and 64-bit operands, respectively. In contrast, ZKPU composes 64-bit processing elements into modular multipliers of arbitrary widths, maintaining over 90\% utilization across all supported operand sizes. These results validate that the proposed upward scalability effectively eliminates the utilization loss inherent to fixed-bitwidth architectures.

\noindent\textbf{On-chip Data Movement.}
\Cref{fig:movement} presents an ablation study of on-chip operand movement per modular multiplication. Starting from a conventional CIOS implementation, adopting FIPS alone slightly increases data movement because of its irregular access pattern. A two-dimensional systolic organization substantially reduces operand transfers through local data reuse, while the proposed twin-PE organization further minimizes communication by sharing operands between multiplication and Montgomery reduction. Overall, the complete ZKPU reduces on-chip operand movement by over an order of magnitude compared with conventional operand-scanning architectures for both 256-bit and 768-bit modular multiplication, demonstrating the effectiveness of the co-designed TSF and Deque buffer.

\noindent \textbf{Discussion.}
Unlike prior accelerators that target either curve-based or hash-based ZKP protocols, ZKPU demonstrates competitive hardware efficiency across both domains. For Groth16 workloads, ZKPU achieves an average 9.4x over PipeZK and 2.6x over LegoZK, while for representative Plonky2 workloads it achieves a consistent 9.5x speedup over UniZK. These results indicate that the proposed architectural abstractions effectively bridge the distinct computational characteristics of curve-based and hash-based proof systems, enabling a single hardware framework to efficiently support both protocol families.
\section{Related Work} \label{sec:related}
Recent advances in ZKP systems have driven the development of dedicated implementation and specialized hardware accelerators targeting the dominant proving kernels. For curve-based ZKPs, GPU-based work such as cuZK~\cite{cuzk}, ElasticMSM~\cite{elasticmsm}, GZKP~\cite{gzkp}, primarily focus on accelerating the key kernels including NTT and MSM by exploiting the massive parallelism of GPUs to improve proving throughput, while numerous FPGA-based accelerators, including SAM~\cite{sam}, BSTMSM~\cite{bstmsm}, CycloneMSM~\cite{cyclonemsm}, MSMAC~\cite{msmac}, PipeMSM~\cite{pipemsm}, targeting specific kernels for  new scheduling strategy and new dataflow to achieve lowest latency in deployment and some in In-Memory Computing~\cite{modsram,mentt,bpntt}. Recent ASICs mentioned extensive earlier for comparison such as PipeZK~\cite{pipezk}, LegoZK~\cite{legozk}, extends this direction to a holistic view by integrating NTT, MSM, Montgomery multiplication, and elliptic-curve arithmetic into an end-to-end ASIC architecture. To its own field, NoCap~\cite{nocap} provides a solution by utilizing hardware-software co-design of the whole ZKP protocols.

The emergence of transparent proof systems, such as Plonky2 and STARKs, has shifted the computational bottlenecks toward hash-based primitives, including Poseidon hashing, Merkle tree construction, and FRI protocols. Consequently, UniZK~\cite{unizk} proposes a programmable accelerator optimized for these kernels. Although these accelerators significantly improve the performance of their target protocols, existing designs remain largely optimized for either curve-based or hash-based ZKPs. As modern proving systems increasingly employ heterogeneous cryptographic primitives, there is a growing need for unified accelerator architectures that can efficiently support both computation models while maximizing hardware utilization and scalability.
\section{Conclusion} \label{sec:conclusion}
This work presented ZKPU, a general-purpose processor architecture that bridged the computational gap between curve-based and hash-based ZKP protocols. By introducing the Finely Integrated Product Scanning (FIPS) Montgomery multiplier with the proposed Twin Systolic Fabric (TSF), together with the Deque buffer memory preprocessing scheme and ZK Engine, the architecture efficiently supported the diverse cryptographic core kernels required across modern ZKP systems. Experimental evaluation demonstrated performance improvements over state-of-the-art ZKP accelerators while simultaneously supporting representative protocols from both families, including Groth16 and Plonky2, eliminating the need for protocol-specific hardware, a capability not offered by existing designs. These results showed that a unified hardware architecture can achieve high performance without sacrificing protocol flexibility, providing a practical foundation for future ZKP systems as the ecosystem continues to evolve.


\bibliographystyle{abbrvnat}
\bibliography{misc/refs}

@article{naf,
  title={Computing multi-scalar multiplication on memory-constrained devices},
  author={No{\"e}l, L{\'e}o and Plantard, Thomas},
  journal={Journal of Cryptographic Engineering},
  volume={16},
  number={2},
  pages={7},
  year={2026},
  publisher={Springer}
}

@inproceedings{rollup,
  title={Analyzing and benchmarking zk-rollups},
  author={Chaliasos, Stefanos and Reif, Itamar and Torralba-Agell, Adri{\`a} and Ernstberger, Jens and Kattis, Assimakis and Livshits, Benjamin},
  booktitle={6th Conference on Advances in Financial Technologies (AFT 2024)},
  pages={6--1},
  year={2024},
  organization={Schloss Dagstuhl--Leibniz-Zentrum f{\"u}r Informatik}
}

@inproceedings{zksnark4,
  title={Sonic: Zero-knowledge SNARKs from linear-size universal and updatable structured reference strings},
  author={Maller, Mary and Bowe, Sean and Kohlweiss, Markulf and Meiklejohn, Sarah},
  booktitle={Proceedings of the 2019 ACM SIGSAC conference on computer and communications security},
  pages={2111--2128},
  year={2019}
}

@inproceedings{zkcnn,
  title={Zkcnn: Zero knowledge proofs for convolutional neural network predictions and accuracy},
  author={Liu, Tianyi and Xie, Xiang and Zhang, Yupeng},
  booktitle={Proceedings of the 2021 ACM SIGSAC Conference on Computer and Communications Security},
  pages={2968--2985},
  year={2021}
}

@inproceedings{zkml,
  title={Zkml: An optimizing system for ml inference in zero-knowledge proofs},
  author={Chen, Bing-Jyue and Waiwitlikhit, Suppakit and Stoica, Ion and Kang, Daniel},
  booktitle={Proceedings of the Nineteenth European Conference on Computer Systems},
  pages={560--574},
  year={2024}
}

@inproceedings{mixed,
  title={Efficient elliptic curve exponentiation using mixed coordinates},
  author={Cohen, Henri and Miyaji, Atsuko and Ono, Takatoshi},
  booktitle={International Conference on the Theory and Application of Cryptology and Information Security},
  pages={51--65},
  year={1998},
  organization={Springer}
}

@inproceedings{deepscaletool,
  title={Deepscaletool: A tool for the accurate estimation of technology scaling in the deep-submicron era},
  author={Sarangi, Satyabrata and Baas, Bevan},
  booktitle={2021 IEEE International Symposium on Circuits and Systems (ISCAS)},
  pages={1--5},
  year={2021},
  organization={Ieee}
}

@article{flalm,
  title={FLALM: A flexible low area-latency Montgomery modular multiplication on FPGA},
  author={Xie, Yujun and Liu, Yuan and Zheng, Xin and Lan, Bohan and Lei, Dengyun and Xiang, Dehao and Cai, Shuting and Xiong, Xiaoming},
  journal={IEEE Transactions on Computers},
  volume={74},
  number={1},
  pages={29--42},
  year={2024},
  publisher={IEEE}
}

@article{scanning2,
  title={A scalable and systolic architectures of montgomery modular multiplication for public key cryptosystems based on dsps},
  author={Mrabet, Amine and El-Mrabet, Nadia and Lashermes, Ronan and Rigaud, Jean-Baptiste and Bouallegue, Belgacem and Mesnager, Sihem and Machhout, Mohsen},
  journal={Journal of Hardware and Systems Security},
  volume={1},
  number={3},
  pages={219--236},
  year={2017},
  publisher={Springer}
}

@article{mentt,
  title={MeNTT: A compact and efficient processing-in-memory number theoretic transform (NTT) accelerator},
  author={Li, Dai and Pakala, Akhil and Yang, Kaiyuan},
  journal={IEEE Transactions on Very Large Scale Integration (VLSI) Systems},
  volume={30},
  number={5},
  pages={579--588},
  year={2022},
  publisher={IEEE}
}

@inproceedings{bpntt,
  title={Bp-ntt: Fast and compact in-sram number theoretic transform with bit-parallel modular multiplication},
  author={Zhang, Jingyao and Imani, Mohsen and Sadredini, Elaheh},
  booktitle={2023 60th ACM/IEEE Design Automation Conference (DAC)},
  pages={1--6},
  year={2023},
  organization={IEEE}
}

@inproceedings{modsram,
  title={ModSRAM: Algorithm-Hardware Co-Design for Large Number Modular Multiplication in SRAM},
  author={Ku, Jonathan Hao-Cheng and Zhang, Junyao and Shan, Haoxuan and Samudrala, Saichand and Wu, Jiawen and Zheng, Qilin and Li, Ziru and Rajendran, Jeyavijayan and Chen, Yiran},
  booktitle={Proceedings of the 61st ACM/IEEE Design Automation Conference},
  pages={1--6},
  year={2024}
}

@article{zcash,
  title={Zerocash: Decentralized anonymous payments from bitcoin},
  author={Ben-Sasson, Eli and Chiesa, Alessandro and Garman, Christina and Green, Matthew and Miers, Ian and Tromer, Eran and Virza, Madars},
  journal={Cryptology ePrint Archive},
  year={2014}
}

@inproceedings{zcoin,
  title={Zerocoin: Anonymous distributed e-cash from bitcoin},
  author={Miers, Ian and Garman, Christina and Green, Matthew and Rubin, Aviel D},
  booktitle={2013 IEEE symposium on security and privacy},
  pages={397--411},
  year={2013},
  organization={IEEE}
}

@article{survey,
  title={Zero-Knowledge Proofs For Privacy-Preserving Systems: A Survey Across Blockchain, Identity, And Beyond},
  author={Gupta, Sandeep},
  journal={Engineering and Technology Journal},
  volume={10},
  number={07},
  pages={5755--5761},
  year={2025}
}

@inproceedings{vml,
  title={Zero knowledge proofs for decision tree predictions and accuracy},
  author={Zhang, Jiaheng and Fang, Zhiyong and Zhang, Yupeng and Song, Dawn},
  booktitle={Proceedings of the 2020 ACM SIGSAC Conference on Computer and Communications Security},
  pages={2039--2053},
  year={2020}
}

@inproceedings{auction,
  title={Verifiable sealed-bid auction on the ethereum blockchain},
  author={Galal, Hisham S and Youssef, Amr M},
  booktitle={International Conference on financial cryptography and data security},
  pages={265--278},
  year={2018},
  organization={Springer}
}

@article{mont,
  title={Modular multiplication without trial division},
  author={Montgomery, Peter L},
  journal={Mathematics of computation},
  volume={44},
  number={170},
  pages={519--521},
  year={1985}
}

@inproceedings{zksnark1,
  title={Succinct non-interactive zero knowledge arguments from span programs and linear error-correcting codes},
  author={Lipmaa, Helger},
  booktitle={International Conference on the Theory and Application of Cryptology and Information Security},
  pages={41--60},
  year={2013},
  organization={Springer}
}

@article{zksnark2,
  title={Pinocchio: Nearly practical verifiable computation},
  author={Parno, Bryan and Howell, Jon and Gentry, Craig and Raykova, Mariana},
  journal={Communications of the ACM},
  volume={59},
  number={2},
  pages={103--112},
  year={2016},
  publisher={ACM New York, NY, USA}
}

@article{zksnark3,
  title={Marlin: Preprocessing zkSNARKs with universal and updatable SRS},
  author={Chiesa, Alessandro and Hu, Yuncong and Maller, Mary and Mishra, Pratyush and Vesely, Psi and Ward, Nicholas},
  journal={Cryptology ePrint Archive},
  year={2019}
}

@inproceedings{zkstark1,
  title={Bulletproofs: Short proofs for confidential transactions and more},
  author={B{\"u}nz, Benedikt and Bootle, Jonathan and Boneh, Dan and Poelstra, Andrew and Wuille, Pieter and Maxwell, Greg},
  booktitle={2018 IEEE symposium on security and privacy (SP)},
  pages={315--334},
  year={2018},
  organization={IEEE}
}

@article{zkstark2,
  title={Scalable, transparent, and post-quantum secure computational integrity},
  author={Ben-Sasson, Eli and Bentov, Iddo and Horesh, Yinon and Riabzev, Michael},
  journal={Cryptology ePrint Archive},
  year={2018}
}

@inproceedings{zkstark3,
  title={Fast reed-solomon interactive oracle proofs of proximity},
  author={Ben-Sasson, Eli and Bentov, Iddo and Horesh, Yinon and Riabzev, Michael},
  booktitle={45th international colloquium on automata, languages, and programming (icalp 2018)},
  pages={14--1},
  year={2018},
  organization={Schloss Dagstuhl--Leibniz-Zentrum fuer Informatik}
}

@incollection{zkp,
  title={The knowledge complexity of interactive proof-systems},
  author={Goldwasser, Shafi and Micali, Silvio and Rackoff, Chales},
  booktitle={Providing sound foundations for cryptography: On the work of shafi goldwasser and silvio micali},
  pages={203--225},
  year={2019}
}

@inproceedings{groth16,
  title={On the size of pairing-based non-interactive arguments},
  author={Groth, Jens},
  booktitle={Annual international conference on the theory and applications of cryptographic techniques},
  pages={305--326},
  year={2016},
  organization={Springer}
}

@article{plonk,
  title={Plonk: Permutations over lagrange-bases for oecumenical noninteractive arguments of knowledge},
  author={Gabizon, Ariel and Williamson, Zachary J and Ciobotaru, Oana},
  journal={Cryptology ePrint Archive},
  year={2019}
}

@techreport{plonky2,
  author      = {Polygon Zero Team},
  title       = {Plonky2: Fast Recursive Arguments with PLONK and FRI},
  institution = {Polygon Labs},
  year        = {2022},
  url         = {https://github.com/0xPolygonZero/plonky2/blob/main/plonky2/plonky2.pdf}
}

@software{starky,
  author = {Polygon Zero},
  title = {Starky: Fast STARK Proving System},
  url = {https://github.com/0xPolygonZero/plonky2/tree/main/starky},
  year = {2022}
}

@inproceedings{poseidon,
  title={Poseidon: A new hash function for $\{$Zero-Knowledge$\}$ proof systems},
  author={Grassi, Lorenzo and Khovratovich, Dmitry and Rechberger, Christian and Roy, Arnab and Schofnegger, Markus},
  booktitle={30th USENIX Security Symposium (USENIX Security 21)},
  pages={519--535},
  year={2021}
}

@article{scanning,
  title={Analyzing and comparing Montgomery multiplication algorithms},
  author={Koc, C Kaya and Acar, Tolga and Kaliski, Burton S},
  journal={IEEE micro},
  volume={16},
  number={3},
  pages={26--33},
  year={1996},
  publisher={IEEE}
}

@inproceedings{pipezk,
  title={Pipezk: Accelerating zero-knowledge proof with a pipelined architecture},
  author={Zhang, Ye and Wang, Shuo and Zhang, Xian and Dong, Jiangbin and Mao, Xingzhong and Long, Fan and Wang, Cong and Zhou, Dong and Gao, Mingyu and Sun, Guangyu},
  booktitle={2021 ACM/IEEE 48th Annual International Symposium on Computer Architecture (ISCA)},
  pages={416--428},
  year={2021},
  organization={IEEE}
}

@inproceedings{legozk,
  title={Legozk: A dynamically reconfigurable accelerator for zero-knowledge proof},
  author={Yang, Zhengbang and Zhao, Lutan and Li, Peinan and Liu, Han and Li, Kai and Zhao, Boyan and Meng, Dan and Hou, Rui},
  booktitle={2025 IEEE International Symposium on High Performance Computer Architecture (HPCA)},
  pages={113--126},
  year={2025},
  organization={IEEE}
}

@article{cuzk,
  title={Cuzk: Accelerating zero-knowledge proof with a faster parallel multi-scalar multiplication algorithm on gpus},
  author={Lu, Tao and Wei, Chengkun and Yu, Ruijing and Chen, Chaochao and Fang, Wenjing and Wang, Lei and Wang, Zeke and Chen, Wenzhi},
  journal={IACR Transactions on Cryptographic Hardware and Embedded Systems},
  volume={2023},
  number={3},
  pages={194--220},
  year={2023}
}

@inproceedings{gzkp,
  title={Gzkp: A gpu accelerated zero-knowledge proof system},
  author={Ma, Weiliang and Xiong, Qian and Shi, Xuanhua and Ma, Xiaosong and Jin, Hai and Kuang, Haozhao and Gao, Mingyu and Zhang, Ye and Shen, Haichen and Hu, Weifang},
  booktitle={Proceedings of the 28th ACM International Conference on Architectural Support for Programming Languages and Operating Systems, Volume 2},
  pages={340--353},
  year={2023}
}

@inproceedings{sam,
  title={Sam: A scalable accelerator for number theoretic transform using multi-dimensional decomposition},
  author={Wang, Cheng and Gao, Mingyu},
  booktitle={2023 IEEE/ACM International Conference on Computer Aided Design (ICCAD)},
  pages={1--9},
  year={2023},
  organization={IEEE}
}

@inproceedings{bstmsm,
  title={Bstmsm: A high-performance fpga-based multi-scalar multiplication hardware accelerator},
  author={Zhao, Baoze and Huang, Wenjin and Li, Tianrui and Huang, Yihua},
  booktitle={2023 International Conference on Field Programmable Technology (ICFPT)},
  pages={35--43},
  year={2023},
  organization={IEEE}
}

@article{cyclonemsm,
  title={Fpga acceleration of multi-scalar multiplication: Cyclonemsm},
  author={Aasaraai, Kaveh and Beaver, Don and Cesena, Emanuele and Maganti, Rahul and Stalder, Nicolas and Varela, Javier},
  journal={Cryptology ePrint Archive},
  year={2022}
}

@article{elasticmsm,
  title={Elastic MSM: A fast, elastic and modular preprocessing technique for multi-scalar multiplication algorithm on GPUs},
  author={Zhu, Xudong and He, Haoqi and Yang, Zhengbang and Deng, Yi and Zhao, Lutan and Hou, Rui},
  journal={IACR Transactions on Cryptographic Hardware and Embedded Systems},
  volume={2024},
  number={4},
  pages={258--284},
  year={2024}
}

@inproceedings{msmac,
  title={MSMAC: Accelerating multi-scalar multiplication for zero-knowledge proof},
  author={Qiu, Pengcheng and Wu, Guiming and Chu, Tingqiang and Wei, Changzheng and Luo, Runzhou and Yan, Ying and Wang, Wei and Zhang, Hui},
  booktitle={Proceedings of the 61st ACM/IEEE Design Automation Conference},
  pages={1--6},
  year={2024}
}

@inproceedings{nocap,
  title={Accelerating zero-knowledge proofs through hardware-algorithm co-design},
  author={Samardzic, Nikola and Langowski, Simon and Devadas, Srinivas and Sanchez, Daniel},
  booktitle={2024 57th IEEE/ACM International Symposium on Microarchitecture (MICRO)},
  pages={366--379},
  year={2024},
  organization={IEEE}
}

@inproceedings{unizk,
  title={Unizk: Accelerating zero-knowledge proof with unified hardware and flexible kernel mapping},
  author={Wang, Cheng and Gao, Mingyu},
  booktitle={Proceedings of the 30th ACM International Conference on Architectural Support for Programming Languages and Operating Systems, Volume 1},
  pages={1101--1117},
  year={2025}
}

@INPROCEEDINGS{pippenger,
  author={Pippenger, Nicholas},
  booktitle={17th Annual Symposium on Foundations of Computer Science (sfcs 1976)}, 
  title={On the evaluation of powers and related problems}, 
  year={1976},
  volume={},
  number={},
  pages={258-263},
  doi={10.1109/SFCS.1976.21}}

@misc{pipemsm,
      author = {Charles. F. Xavier},
      title = {{PipeMSM}: Hardware Acceleration for Multi-Scalar Multiplication},
      howpublished = {Cryptology {ePrint} Archive, Paper 2022/999},
      year = {2022},
      url = {https://eprint.iacr.org/2022/999}
}


\end{document}